# Evidence for multiple gaps in specific heat of LiFeAs crystals

F. Wei,[1] F. Chen,[1] K. Sasmal,[1] B. Lv,[2] Z. J. Tang,[2] Y. Y. Xue,[1] A. M. Guloy,[2] and C. W. Chu[1,3,4]

[1]Department of Physics and TCSUH, University of Houston, Houston, Texas 77204-5002, USA

[2]Department of Chemistry and TCSUH, University of Houston, Houston, Texas 77204-5002, USA

[3]Lawrence Berkeley National Laboratory, 1 Cyclotron Road, Berkeley, California 94720, USA

[4]Hong Kong University of Science and Technology, Hong Kong, China

**Abstract**

The zero-field specific heat of LiFeAs was measured on several single crystals selected from a bulk sample. A sharp $\Delta C_p/T_c$ anomaly $\approx$ 20 mJ/mole·K$^2$ was observed. The value appears to be between those of SmFeAs(O$_{0.9}$F$_{0.1}$) and (Ba$_{0.6}$K$_{0.4}$)Fe$_2$As$_2$, but bears no clear correlation with their Sommerfeld coefficients. The electronic specific heat below $T_c$ further reveals a two-gap structure with the narrower one only on the order of 0.7 meV. While the results are in rough agreement with the $H_{c1}(T)$ previously reported on both LiFeAs and (Ba$_{0.6}$K$_{0.4}$)Fe$_2$As$_2$, they are different from the published specific-heat data of a (Ba$_{0.6}$K$_{0.4}$)Fe$_2$As$_2$ single crystal.

Intense research activity was recently stimulated by the discovery of superconductivity in the FeAs-based superconductors. These compounds are often compared with the well investigated cuprates. In particular, the hope that both might share similar pairing mechanisms has been raised after observations of the existence of spin-density waves in the FeAs-based compounds [1-3]. Differences between these two families were soon discovered, and a rough picture of the FeAs-based superconductors, *i.e.* the superconductivity might be associated with dynamic interband spin-coupling, has emerged [4,5]. However, many questions still remain. Although a more-or-less universal trend is expected based on the common layered structure and the comparable superconducting transition temperatures $T_c$ within the FeAs family, the properties observed (even the basic thermodynamic parameters) are rather divergent. The doping dependencies of $T_c$, the Sommerfeld coefficient γ (a measure of the density of states) [4,6], the specific-heat anomaly $\Delta C_p/T$ at $T_c$ (a parameter representing the pair-coupling strength), the temperature dependence of the superconducting gap below $T_c$ (*i.e.* the wave-function symmetry) [7-10] and the residual $C_p/T$ at the zero temperature limit (an indicator of the possible phase-separation) all vary significantly from one member to another. Although both RFeAs($O_{0.9}F_{0.1}$) and ($Ba_{0.6}K_{0.4}$)$Fe_2As_2$ are near the optimum doping level and have comparable $T_c$, for example, the specific-heat anomaly $\Delta C_p/T$ around $T_c$ is ten times as high in ($Ba_{0.6}K_{0.4}$)$Fe_2As_2$, where R is a rare earth element [6]. A much lower γ ≈ 6-8 mJ/mole·K$^2$ in LaFeAs($O_{0.9}F_{0.1}$) was then used to accommodate the difference. However, an apparent γ = 121 mJ/mole·K$^2$, but a $\Delta C_p/T_c$ as small as that of LaFeAs($O_{0.9}F_{0.1}$), is observed in SmFeAs($O_{0.9}F_{0.1}$) [6,7]. The γ remains above 50 mJ/mole·K$^2$ even after corrections for possible Schottky-like anomalies [7]. In addition, some of the reports of electronic specific heat, $C_{p,e}/T$, below $T_c$ have been so different that *d*-wave and *s*-wave pairings were proposed for LaFeAs($O_{0.9}F_{0.1}$) and (Ba$_{0.6}$K$_{0.4}$)Fe$_2$As$_2$, respectively [6,10]. The situation is actually even more complicated. Different gap features have been suggested for the same (Ba$_{0.6}$K$_{0.4}$)Fe$_2$As$_2$ crystals from two different bulk probes, specific heat and the lower critical field $H_{c1}$ [6,11]. Although several surface probes, such as angle-resolved photo-emission spectroscopy (ARPES), reveal similar two-gap characteristics, the reported gap values are also rather different [12,13]. It is unclear whether such divergence reflects the intrinsic compound-to-compound variation, sample quality, or nature of the probes. Here we report the zero-field specific-heat observed for LiFeAs single crystals. A sharp $\Delta C_p/T_c$ anomaly ≈ 20 mJ/mole·K$^2$ together with significant residual $C_{p,e}/T$ down to 2 K was observed. The deduced $\Delta C_p/\gamma T_c$ is different from those for SmFeAs($O_{0.9}F_{0.1}$) and (Ba$_{0.6}$K$_{0.4}$)Fe$_2$As$_2$, suggesting a coupling strength varying from one compound to another within the FeAs family. The data, however, can be fitted well with a two-gap *s*-wave structure with the lower gap around 0.7 meV, in line with both our $H_{c1}$ data on similar LiFeAs crystals [14] and the $H_{c1}$ and some ARPES data of (Ba$_{0.6}$K$_{0.4}$)Fe$_2$As$_2$ [11,13].

Bulk LiFeAs samples were synthesized from high temperature reactions of high purity Li, Fe, and As, as previously reported [15]. The X-ray diffraction of the polycrystalline samples indicates single phase, corresponding to the LiFeAs structure (Fig. 1a, with Cu K$_\alpha$ line). Superconductivity of the LiFeAs sample was verified using a 5 T Quantum Design SQUID magnetometer (Fig. 1b). The 10 Oe zero-field-cooled and field-cooled magnetizations reveal a bulk superconducting transition with an onset around 17 K. Five grains with shiny cleavage surfaces and in-layer dimensions around 0.1 mm were selected from the samples. Relevant properties of both initial powder and the selected single crystal assembly were measured using a Quantum Design Physical Properties Measurement System (PPMS) over the temperature range between 1.8 and 160 K. The crystals were placed with their *ab* planes along the sample platform. Apiezon N-grease was used to ensure good sample contact.

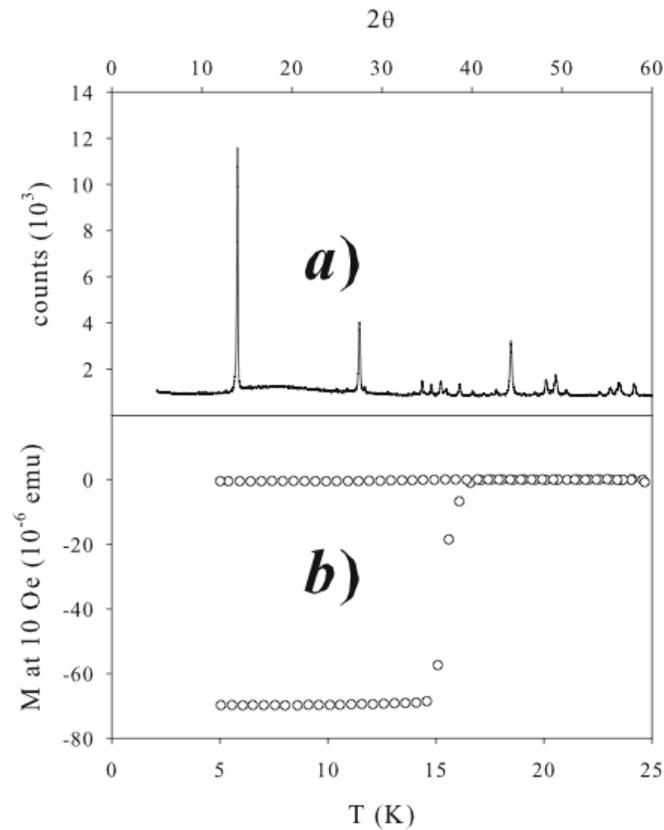

Figure 1. a) XRD pattern of LiFeAs bulk sample that exhibits preferred orientation along (001). The data were collected with Cu $K_\alpha$ lines. The small impurity peaks belong to $FeAs_2$. b) Magnetization of similar LiFeAs single crystals selected from the bulk samples.

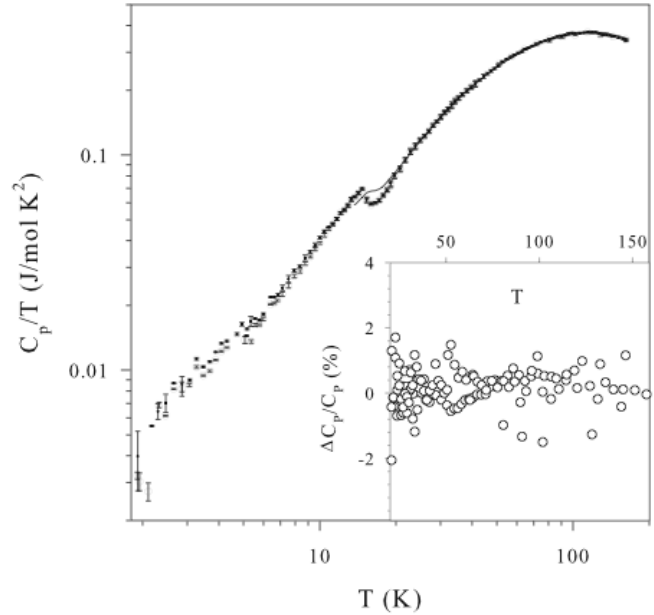

Figure 2. The reproducibility of the data with error bars representing the standard deviations of 4-8 sequential repeat measurements. The small dots and large open circles are the data of the first run and the second run (one day later) on the crystals, respectively. The solid line represents the data of a large piece of the parent sample. Inset: The deviations of the above individual data points from their smoothed average.

The small crystal size available presents an experimental challenge. It is widely accepted that the sample contribution to $C_p$ should be more than 1/3 of the total $C_p$ to achieve accuracy better than 1% [16]. Our total crystal-mass ≈ 0.5 mg, unfortunately, accounts for only 10-20% of the platform $C_p$ over the temperature range explored. Two adverse effects, therefore, may result: larger random noise due to the platform background, as well as a systematic distortion due to the inability to properly model the sample-to-platform thermal retardation. To verify the data reliability, both the uncertainty $\delta C_p$ given by the PPMS software and the standard-deviation $\delta_1C_p$ from 4-8 consecutively repeated measurements were monitored. The $\delta_1C_p$ appears to be much higher than $\delta C_p$, and is presented as error bars in the discussion below (*e.g.* Fig. 2). Such fluctuation, however, is less than a few percent above 20 K (the difficulties at low-T will be addressed later). A second run one day later further demonstrates good reproducibility (Fig. 2). To explore the possible system distortion, the data of the bulk piece, which accounts for more than half of the total $C_p$ observed, were analyzed. The above three data sets are in good agreement, except for the much broader transition and the lower anomaly in the bulk data. To make the agreement more explicit, a "reference" was deduced by smoothing and averaging all data points, and the deviation of individual points was deduced. The differences are only at the 1% level (inset, Fig. 2). Several factors may contribute to this mass-insensitive reproducibility: the good effective sample-to-platform thermal conductance with its retardation time-constant $\tau_2$ being only less than 5% that of the platform-to-system for the bulk sample; the preferred thin-plate crystal-shape with the large area-to-thickness ratio; and the assumed good contact between the flat *ab* surfaces and the platform. It should be pointed out, however, that the data points below 2.2 K appear to be suspicious with large scattering and will only be treated as a tentative reference below.

It is interesting to note that a Debye approximation of $C_p/T \propto \gamma + \beta T^2$ cannot simultaneously satisfy the $C_p$ and the associated entropy S even if the temperature window is extremely narrow, *e.g.* 18-22 K, and that a $T^5$ term has to be added. Similar non-Debye behaviors have been noticed previously. A six-term polynomial together with a prefixed γ was used to fit the data of a $Ba_{0.6}K_{0.4}Fe_2As_2$ crystal between 35 and 50 K [6]. It should be pointed out that even such higher order expansion leads to un-physical negative values above 60 K. In particular, the most crucial γ has to be preset due to both the narrow temperature range available and the large number of free parameters invoked. Consequentially, a scaling model and an estimated upper critical field $H_{c2}$ = 100 T at T = 0 based on the Werthamer-Helfand-Hohenberg (WHH) relation were used to determine the γ [6]. Previous data, however, have already demonstrated that the WHH relation significantly underestimates the zero-temperature $H_{c2}$ value in the case of $LaFeAs(O_{0.9}F_{0.1})$ [17]. A reliable estimation of the phonon background, therefore, appears to be the key for the analysis of specific heat data. Tropeano *et al.* and Baker *et al.*, fortunately, have demonstrated that such a non-Debye trend may be caused by a large Einstein contribution and a rather low Debye temperature $T_D$ [7,18]. The normal-state specific heat, $C_{p,n}$, was therefore proposed as:

$$C_{p,n} = \gamma T + A_D C_D(T,T_D) + A_E C_E(T,T_E), \quad (1)$$

where $C_D$, $C_E$, $T_E$, $A_D$, and $A_E$ are Debye and Einstein functions, the Einstein temperature, and two fitting parameters, respectively. The $C_p$ of several FeAs-based compounds can be well fit using Eq. 1 [7,18]. Thus, this approximation is adopted here. Both the $C_p$ and the associated entropy $S = \int_0^T \frac{C_p}{T'} dT'$ between 17 and 160 K can be fitted well (solid lines in Fig. 3 and its insert). The fitting parameters γ = 0.019(1) J/mole·K², $T_D$ = 240(4) K, and $T_E$ = 410(5) K are in reasonable agreement with those reported for a LiFeAs pressed powder sample [7]. It should be pointed out that the lower $T_c$ of LiFeAs actually makes the estimation of $\gamma$ (as well as the low-T electronic specific heat, to some degree) much more robust and model-independent. For example, a three-term polynomial fit of $\gamma + \beta_3 T^2 + \beta_5 T^4$ below 30 K leads to a similar value of $\gamma = 0.020$ J/mole·K$^2$. This can be easily understood. The Einstein term contributes noticeably only above 45 K (inset, Fig. 3). The three-term polynomial expansion, therefore, will be a good approximation of Eq. 1 below $0.2 T_D \approx 50$ K. Such good reliability, however, has an even deeper significance in the case of LiFeAs. The $\gamma$ value is actually defined by the entropy constraint of

$$\gamma T_c = \int_0^{T_c} \frac{C_p dT}{T} - S_{phonon},$$

where $S_{phonon}$ is the (model-dependent) phonon contribution. Although the total entropy $\int_0^{T_c} \frac{C_p dT}{T}$ of LiFeAs is 0.5 J/mole·K$^2$ at 16 K > $T_c$, the estimated $S_{phonon} \approx 0.2$ J/mole·K$^2$ accounts only for 40% of the observed value. Possible uncertainties, *e.g.* the phonon-background analysis above 20 K and the possible distortions below 2.2 K, can hardly cause a significant change in $\gamma$.

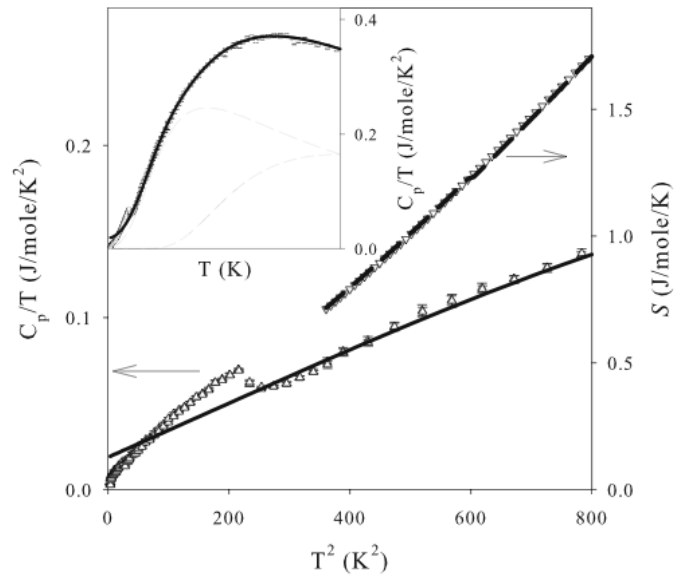

Figure 3. The observed $C_p/T$ (triangles with error bars representing the standard deviations of 4-8 sequential repeat measurements) and entropy S (inverted triangles) of a LiFeAs single crystal. Solid line: the expected normal-state $C_{p,n}/T$ from the fitting. Inset: The $C_p/T$ observed up to 160 K with error bars representing the standard deviations of 4-8 sequential repeat measurements. The thick lines are the fits from Eq. 1. The upper and lower thin dashed lines are the Debye and Einstein contributions, respectively.

The difference, $\Delta C_p/T = (C_p - C_{p,n})/T$, associated with the superconducting transition is consequently deduced (Fig. 4b). Both the thermodynamic $T_c \approx 15.4$ K and the $\Delta C_p/T$ jump $\approx 20$ mJ/mole·K$^2$ at $T_c$ are significantly higher than those reported for the sintered powder LiFeAs sample [7]. The spread of the jump is less than 2 K on the higher temperature side, demonstrating a negligible $T_c$ spread despite the fact that five different crystals were randomly selected from the bulk. It is interesting to note that the observed anomaly is 2-3 times lower than that for Ba$_{0.6}$K$_{0.4}$Fe$_2$As$_2$ $\approx 49$ mJ per Fe·K$^2$. Even the more meaningful parameter $\Delta C_p/(\gamma-\gamma_0)T_c$ is noticeably different between the two compounds, where $\gamma_0$ is the residual $C_p/T$ at the zero-temperature limit. The $\Delta C_p/(\gamma-\gamma_0)T_c \approx 1.2$ for LiFeAs is smaller than the reported value of 1.6 for Ba$_{0.6}$K$_{0.4}$Fe$_2$As$_2$ based on the adopted WHH H$_{c2}$ of 100 T. The possible $\gamma_0$ uncertainty in LiFeAs can only shift the ratio of LiFeAs by ±0.2 (as will be discussed below), and the possible corrections for the WWH approximation may even raise that of Ba$_{0.6}$K$_{0.4}$Fe$_2$As$_2$ to be twice as large. The large compound-to-compound variation of $\Delta C_p/(\gamma-\gamma_0)T_c$ as described above, therefore, occurs again. It should be noted, however, that both LiFeAs and Ba$_{0.6}$K$_{0.4}$Fe$_2$As$_2$ contain no magnetic elements, *i.e.* it is difficult to attribute the divergence to magnetic anomalies alone. It is also interesting to note that this ratio is closely related to both the coupling strength and possible gap structures based on the α model [19]. The supercarrier's contribution in the model is estimated within the BCS framework with the pairing strength $2\Delta/kT_c$ as the main fitting parameter, where $\Delta$ is the superconducting gap at the zero-temperature limit. The different $\Delta C_p/(\gamma-\gamma_0)T_c$s, therefore, imply different coupling strengths $2\Delta/kT_c$s within the same FeAs family.

Returning to the possible uncertainties, it is interesting to note that the $\gamma_0$ appears to be the main factor in our case. Large $\gamma_0$ has been observed in both cuprates and FeAs-based superconductors [7,20]. Contributions from isolated "normal" inclusions have been widely accepted as its origin. The carriers within such inclusions may not take part in superconductive condensation, and lead to a non-zero $\gamma_0$. Extrapolating $\gamma_0$ from a limited temperature window, however, is model-dependent (Fig. 4). In addition to the phonon contributions, electronic contributions also exist. Either a linear term, $aT$, or an exponential term $a \cdot e^{-\Delta/kT}/T^{2.5}$, is expected for the *d*-wave or the *s*-wave pairings, respectively. In the case of LiFeAs, unfortunately, the lowest measurement temperature ≈ 1.8 K corresponds only to a $T/T_c \approx 0.15$, and the doubts about the data below 2.2 K makes the situation even worse. To ignore the electronic contribution under such conditions is doubtful even in the *s*-wave superconductors. The continuous decrease of $\Delta C_p/T$ down to the lowest temperatures observed (Fig. 4) indeed demonstrates a significant electronic contribution. Different models have therefore been tried, but give rather different results. Other factors, *e.g.* the phonon contributions, have also been considered, but affect our conclusion to a much smaller degree.

Fortunately, both upper and lower limits of $\gamma_0$ can still be reliably settled. First of all, one can get an upper limit of $\gamma_0 \approx 6.2$ mJ/mole·K$^2$ by ignoring all electronic contributions below 2.2 K (the dot-dashed line in Fig. 4a). The true $\gamma_0$ value should be significantly lower if a noticeable electronic component exists. On the other hand, a $\gamma_0 = 0$ will be the lowest possible value. This limit is also unlikely to be reached. The $\gamma_0$ of all other FeAs-based compounds is larger than 1 mJ/mole·K$^2$ [2,6,7]. A $\gamma_0$ value in the middle of the two limits, *e.g.* 3 mJ/mole·K$^2$, seems to be more likely. We, therefore, have to explore the situation assuming a true $\gamma_0$ anywhere between 0 and 6.2 mJ/mole·K$^2$.

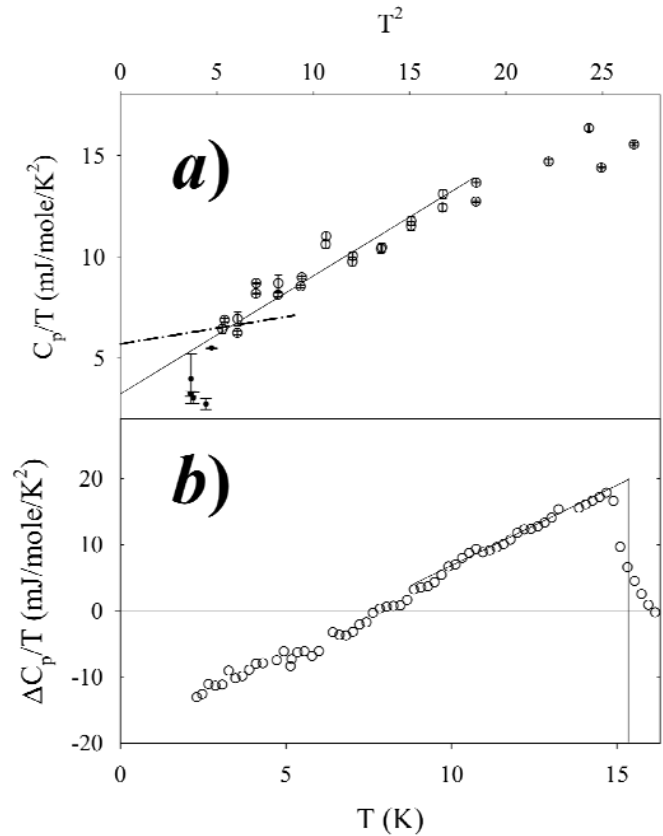

Figure 4. a) The $C_p/T$ observed between 2.2 and 5 K (open circles with error bars representing the standard deviations of 4-8 sequential repeat measurements) and the data below 2.2 K (solid circles with error bars representing the standard deviations of 4-8 sequential repeat measurements), which may be less reliable due to the small crystal mass (see text). The solid line is the linear fit of $\gamma_0 + \beta'_3 T^2$ below 4 K, and the dot-dashed line assumes no electronic contribution below $T_l$. b) The $\Delta C_p/T$ below $T_c$. A thermodynamic $T_c \approx 15.4$ K and a jump $\Delta C_p/T \approx 20$ mJ/mole·K$^2$ around $T_c$ are estimated through linear extrapolations (solid lines).

Calculations based on the $\alpha$ model were carried out for comparison with the observed electronic specific heat $C_{p,e}/T(\gamma-\gamma_0)$ with $\gamma_0$ = 3.5, 0, and 6 mJ/mole·K$^2$ in Figs. 5a, 5b, and 5c, respectively. The normal-state carrier energy, $\varepsilon$, is replaced by a quasiparticle energy, $E = \sqrt{\varepsilon^2 + \Delta^2(t)}$, below $T_c$ in this phenomenological model, where $\Delta$ and $t$ are the superconductive gap and the deduced temperature $T/T_c$, respectively. The associated entropy, S, and heat capacitance, C, of the *s*-wave BCS superconductors can be calculated from

$$\frac{S}{\gamma_n T_c} = -\frac{6\Delta(0)}{\pi^2 k_B T_c} \int_0^\infty [f \ln f + (1-f) \ln(1-f)] dy \qquad (2)$$

and

$$\frac{C_{p,e}}{\gamma_n T_c} = t \frac{d(S/\gamma_n T_c)}{dt}, \qquad (3)$$

with the adjustable parameter $\alpha = 2\Delta(0)/kT_c$ representing the coupling strength, where $f = \dfrac{1}{e^{E/kT}+1}$, $y = \varepsilon/\Delta(0)$, and $C_{p,e}$ is the specific heat of supercarriers. An approximate gap function of $\Delta(t)/\Delta(0) = \tanh[1.837(1/t-1)^{0.51}]$ is also used [21]. In the case of two-gap $s$-wave pairing, the supercarrier contribution $\Delta C_p/T(\gamma-\gamma_0)+1$ will be a simple sum of $[r \cdot C_{p,e}(\alpha_1)+(1-r)\cdot C_{p,e}(\alpha_2)]/T(\gamma-\gamma_0)$, where r, $\alpha_1$, and $\alpha_2$ are the mixing ratio and the coupling strengths of the two gaps, respectively.

The two-gap fit reproduces the data well over all of the possible $\gamma_0$ range despite the noticeable data fluctuation associated with the small crystal mass (solid lines, Fig. 5). The best one-gap fits (dashed lines, Fig. 5), on the other hand, are worse even under the assumed $\gamma_0 = 6.2$ mJ/mole·K$^2$. It should be pointed out that the poorer fit of the one-gap model is a direct result of the large $\Delta C_p/(\gamma-\gamma_0)T_c$ anomaly observed. A strong coupling strength $\alpha_1 > 3.0$ is therefore expected, which demands a negligible low-T tail. The conclusion that a multiple-gap configuration is preferred in LiFeAs, in our view, is not affected by the uncertainty of the $\gamma_0$ value. Unfortunately, the fitting parameters of the two-gap model depend on the adopted $\gamma_0$. The $\alpha_1$ associated with the larger gap is insensitive to the change of $\gamma_0$, as expected. Its value changes slightly from 3.2 at $\gamma_0 = 0$ to 3.6 at $\gamma_0 = 6.2$ mJ/mole·K$^2$. Similarly, the mixing ratio r varies only moderately from 0.7 to 0.85 over the range of $\gamma_0$ values. The smaller coupling strength $\alpha_2$, on the other hand, changes significantly from 0.7 to 1.4 while $\gamma_0$ decreases from 6.2 to 0 mJ/mole·K$^2$. It should also be noted that although the possible $\gamma_0$ range is rather broad, the middle value of 3 mJ/mole·K$^2$ seems to be the most likely case. The associated parameters of $\alpha_1$

= $2\Delta_1/k_BT_c$ = 3.5, $\alpha_2$ = $2\Delta_2/kT_c$ = 1.2, and r = 0.75 therefore suggest a narrow gap, $\Delta_2$, on the order of 0.7 meV. It is also interesting to note that the fitting parameters are in rough agreement with our analysis of the $H_{c1}$ data on two similar LiFeAs crystals, *i.e.* with $\Delta_1$ = (2.7±0.8)$kT_c$, $\Delta_2$ = (0.5±0.2)$kT_c$, and r = 0.5±0.2 [14]. Despite the moderate data resolution and the large $\gamma_0$ uncertainty, the data clearly show that additional low energy gaps (or possible nodes) are needed.

Both the existence of a rather narrow gap and even its strength $2\Delta_1/kT_c$ extracted here are also in rough agreement with the $H_{c1}$ data [11] and recent μSR/ARPES data [13] on similar $Ba_{0.6}K_{0.4}Fe_2As_2$ single crystals. Our data, therefore, strongly suggest that the two-gap pairing with a rather narrow gap, which dominates the low T specific heat, occur in both LiFeAs and $Ba_{0.6}K_{0.4}Fe_2As_2$ that have different crystal structures. The reported $C_{p,e}/T$ data of $Ba_{0.6}K_{0.4}Fe_2As_2$ [6], however, appear to be rather different for reasons still not quite understood. Thus, more specific heat measurements on various FeAs-based compounds are recommended.

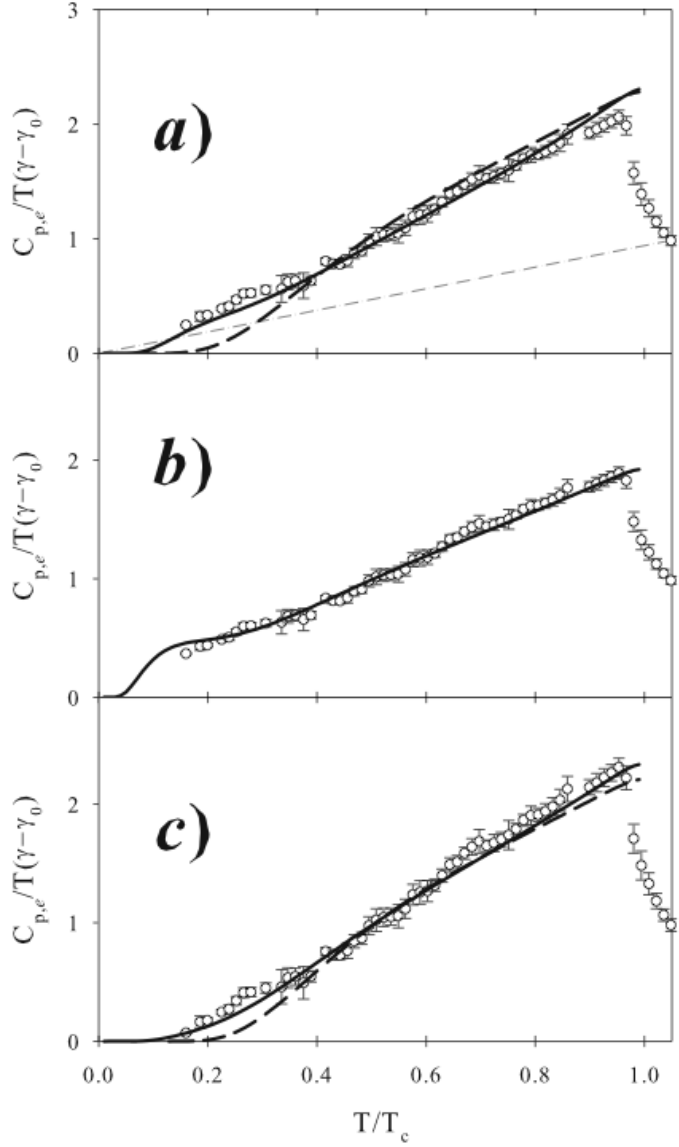

Figure 5. Comparison of the electronic $C_{p,e}/(\gamma-\gamma_0)T$ observed with models. Circles with error bars representing the standard deviations of 4-8 sequential repeat measurements: data; solid lines: the $\alpha$ model fits with two $s$-wave gaps; dashed lines: fits with single $s$-wave gap; dot-dashed line: the expected low T approximation of single $d$-wave gap in Ref. 21. *a*) at $\gamma_0 = 3$ mJ/mole·K$^2$. *b*) at $\gamma_0 = 0$ mJ/mole·K$^2$. *c*) at $\gamma_0 = 6.2$ mJ/mole·K$^2$.

Verifying the possible gap nodes is difficult due to the low $T_c$ of LiFeAs and data resolution. We can therefore only offer some indirect and tentative suggestions. The $C_{p,e}/(\gamma-\gamma_0)T$ of a single-gap $d$-wave superconductor is expected to appear as a linear term $aT$ below $0.3T_c$ with a likely coefficient $a \approx (\gamma-\gamma_0)/T_c$ [22]. This is clearly different from the observed data, with both $\gamma_0 = 0$ and 6.2 mJ/mole·K$^2$ (Figs. 5b and 5c). Even in the case of $\gamma_0 = 3$ mJ/mole·K$^2$ (Fig. 5a), the model calculation is in significant disagreement with the data between 2 and 4 K (dot-dashed line, Fig. 5a). However, as previously noted, the exact $a$ value depends on the detailed angle dependency of the gap around the nodes [23]. The possible multi-gap structures should further complicate the situation. Our data, therefore, cannot rule out the existence of nodal gaps.

A spin-density wave has been proposed as a competing/coexisting excitation against superconductivity in FeAs-based compounds. Experimentally, however, no evidence has been reported in LiFeAs, except for the observations of two anomalies between 40 and 60 K in similar NaFeAs single crystals [24]. The $C_p/T$ of the selected LiFeAs crystals was therefore investigated over the whole temperature range between 2 and 160 K. The differences between the data and the smooth Debye/Einstein fit were integrated to set the upper limit for the possible entropy involved. No deviations associated with an entropy change $|\Delta S| > 0.001R$, the estimated experimental resolution, can be noticed, where R is the Avogadro constant. Compared with the entropy involved in the superconducting transition of LiFeAs around 0.1R and that, $\approx 0.01R$, of the proposed magnetic anomalies in NaFeAs [24], there is no evidence for noticeable static magnetic excitations.

In summary, the specific heat of an assembly of LiFeAs single crystals reveals a multi-gap feature with a small gap of about 0.7 meV dominating the low temperature quasiparticle excitations. A significant contribution from Einstein phonons is observed, as well as a noticeable residual linear term $\gamma_0$.

**Acknowledgments**

The work in Houston is supported in part by the U. S. Air Force Office of Scientific Research, the T. L. L. Temple Foundation, the John J. and Rebecca Moores Endowment, the Robert A. Welch Foundation (under Grant no. E-1297), and the State of Texas through the Texas Center for Superconductivity at the University of Houston; and at Lawrence Berkeley Laboratory by the Director, Office of Science, Office of Basic Energy Sciences, Division of Materials Sciences and Engineering of the U.S. Department of Energy under Contract No. DE-AC03-76SF00098.

Afterword: Following the initial submission of this work, a recent ARPES work on the LiFeAs crystal came to our attention {arXiv:1001.1147v1 [cond-mat.supr-con] (2010)}. It is interesting to note that the multigap structure and the gap widths of 1.5 meV and 2.5 meV over the hole-like and electron-like pockets, respectively, reported in the other work are in rough agreement with the 0.7 meV and 2.5 meV reported here. This further suggests that the mulitiband feature of LiFeAs may naturally lead to a rather complicated gap structure.